\documentclass{cs23proc}

\usepackage{kantlipsum}

\editors{Takeru Suzuki and the Cool Stars 23 Organizing Team}
\publisher{Zenodo}
\conference{The 23th Cambridge Workshop on Cool Stars, Stellar Systems, and the Sun (Cool Stars 23)}
\conferencedate{2026}

\title{The impact of tidal locking on Earth-like planetary dynamos}
\author{J. P. Hidalgo $^{1,2}$ \&
        D. R. G Schleicher $^{1}$}

\affiliation{$^{1}$ Dipartimento di Fisica, Sapienza, Università di Roma, Piazza le Aldo Moro 5, 00185 Roma, Italy \\
			 $^{2}$ INAF, Observatory of Abruzzo, Via Mentore Maggini snc, I-64100 Teramo, Italy \\ }

\shorttitle{The impact of tidal locking on Earth-like planetary dynamos}
\shortauthors{Hidalgo \& Schleicher}

\abs{We investigate how tidal locking affects the planetary dynamo and the magnetospheres of Earth-like planets orbiting the habitable zones of M dwarfs. We couple an empirical stellar wind model with two distinct planetary dynamo paradigms, direct rotational scaling and energy-flux scaling, using a Constant Time Lag (CTL) model to simulate continuous tidal spin-down until synchronization. We model two hypothetical planets based on a highly dissipative Modern Earth and a less dissipative, rapidly rotating Early Earth representative of the Paleoarchean. Our results show that tidal locking acts as a severe negative feedback on planetary magnetism across both paradigms, where the dipolar field frequently collapses before full tidal synchronization is reached. Essentially, a shielded atmospheric environment around M dwarfs requires a host massive enough ($\geq 0.32 M_\odot$) to place the HZ at wider orbital distances, sparing the planet from the sub-Alfvénic environment, and the internal tidal dissipation of the planet must be sufficiently low to prevent rapid rotational decay. For mid-to-late M dwarfs, the inevitable tidal locking and the extreme stellar wind pressures result in a total collapse of the magnetic shield, reducing the likelihood of atmospheric protection to essentially zero across the habitable zone.}

\begin{document}

\maketitle

\section{Introduction}
M dwarfs are the predominant stellar population in our galaxy, accounting for roughly $70-75\%$ of the stars \citep{Bochanski2010}. In M-dwarf systems, the habitable zone (HZ), defined as the region where a terrestrial planet can maintain liquid water on its surface, is located significantly closer to the host star due to their low luminosities. This proximity provides a good contrast to detect exoplanets, via transit and radial velocity methods \citep{Kasting2014,Deeg2018}. However, within these close-in HZs, terrestrial planets are highly susceptible to tidal synchronization, which can increase their rotation periods significantly \citep{Leconte2015, Barnes2017, Childs2022}. 

The intrinsic planetary magnetic field is a key ingredient to protect the atmosphere against the erosion produced by the stellar winds \citep{See2014}. While several studies have explored the effects of the stellar pressure on the magnetospheres \citep[e.g.][]{Lammer2007,Vidotto2013,Hidalgo2025b}, one important effect that is frequently overlooked is the fundamental dependence on rotation of the planetary dynamo. Some studies have explored the impact of tidal locking using rotation-dependent scaling, which reduces the magnetic dipolar moments, and consequently the magnetospheric sizes, significantly \citep{Griessmeier2004, Griessmeier2005, Lammer2007, Griessmeier2009}. Updated models, such as those from \cite{Zuluaga2013} and \cite{RodriguezMozos2019} have included updated dynamo scaling laws. However, these models do not account for the gradual spin-down of the planet towards synchronous rotation, and they usually rely directly on fixed-rotation criteria or linear tidal-locking timescales.

In this work, we summarize the model presented by \cite{Hidalgo2026}, which corresponds to a self-consistent framework that couples empirical stellar winds, a gradual spin-down towards synchronization, and the planetary magnetic field with two distinct dynamo scaling laws. In Section~\ref{framework} we describe the used model. In Section~\ref{results} we apply our framework to a sample of 15 M dwarfs, and our conclusions are summarized in Section~\ref{conclusions}.

\section{Theoretical framework \label{framework}}
In this section, we describe our approach to model tidal locking, the adopted dynamo scaling laws, and our assumptions to estimate the size of the upstream magnetosphere of planets orbiting M dwarfs. A detailed description of this model is presented in  \cite{Hidalgo2026}. 

\subsection{Rotational evolution}
To ensure that the planet smoothly spins down until synchronization is reached, we solve the angular evolution equation from the classic Constant Time Lag (CTL) models \citep{Barnes2017}. Assuming zero obliquity and zero eccentricity, it reduces to
\begin{equation}
    \frac{d\omega}{dt} = \frac{3 k_2}{\alpha M_\mathrm{pl}} \frac{G^2 M_\star^2 (M_\star + M_\mathrm{pl}) r_\mathrm{pl}^3}{n r_\mathrm{orb}^6} \tau \left( 1 - \frac{\omega}{n} \right),
\end{equation}
where $\omega$ is the planetary rotation rate, $k_2$ is the second-degree Love number, $\alpha$ is the dimensionless moment of inertia, $M_\mathrm{pl}$ is the planetary mass, $G$ is the gravitational constant, $M_\star$ is the mass of the host star, $n$ is the mean orbital motion, $r_\mathrm{orb}$ is the orbital distance, $r_\mathrm{pl}$ is the planetary radius and $\tau$ is the tidal time lag. Assuming constant planetary parameters and a time-independent orbit, the solution is given by
\begin{equation}
        \omega(t) = n + (\omega_\mathrm{i} -n ) \exp\left( - \frac{t}{\tau_\mathrm{lock}}  \right),
\end{equation}
where $\omega_\mathrm{i}$ is the initial rotation rate, and
\begin{equation}
    \tau_\mathrm{lock} = \frac{1}{\tau} 
    \frac{\alpha}{3 k_2} \frac{M_\mathrm{pl} r_\mathrm{orb}^6}{G M_\star^2 r_\mathrm{pl}^3}
\end{equation}
is the characteristic tidal synchronization timescale.
\subsection{Planetary dynamo scaling}
For the planetary magnetic field $B_\mathrm{pl}$, we assume a dipolar configuration, based on the large-scale magnetic field of Earth, i.e. \citep{Vidotto2013}
\begin{equation}
        B_\mathrm{pl}(r) = \frac{1}{2} B_\mathrm{p,0} \left( \frac{r_\mathrm{pl}}{r}\right)^3,
\end{equation}
where $B_\mathrm{p,0}$ is the magnetic field at the pole.

To quantify how the spin-down affects the magnetic field of the planet, we use the following models based on dynamo scaling laws:
\subsubsection{Model A: Direct rotation scaling}
The dipolar magnetic field scales directly with rotation, which is
\begin{equation}
    B_\mathrm{p,0}^{(\omega)}(t) = \frac{\omega(t)}{\omega_\mathrm{i}} B_\mathrm{p,0},
\end{equation}

\subsubsection{Model B: Energy-flux scaling }
The magnetic field strength is primarily driven by the convective heat flux, and rotation plays a critical role in shaping its geometry \citep{Christensen2006}. A key diagnostic of the magnetic topologies of rotating dynamos is the local Rossby number, which scales as $\mathrm{Ro}_\ell \propto \omega^{-7/6}$ \citep{Olson2006}.  In regimes where Coriolis forces dominate $(\mathrm{Ro}_\ell \leq 0.1)$, stable dipolar solutions emerge, whereas higher inertial contributions $(\mathrm{Ro}_\ell \geq 0.1)$ lead to multipolar configurations \citep{Gastine2012}. Numerical simulations show that the boundary between these two regimes is typically inherently sharp \citep[e.g.][]{Christensen2006}. Therefore, we define
\begin{equation}
    B_\mathrm{p,0}^{(\omega)}(t) = \left[ f_\mathrm{dip,TL} + \frac{f_\mathrm{dip} - f_\mathrm{dip,TL}}{1 + e^{k(\mathrm{Ro}_\ell(t) - \mathrm{Ro}_\mathrm{\ell,crit})}} \right] B_\mathrm{p,0},
\end{equation}
where $f_\mathrm{dip} = 1$ is the initial dipolarity, $k$ is a steepness parameter, and $f_\mathrm{dip,TL}$ is the dipolar fraction after the topology transition. As the strength of the dipolar component typically differs by an order of magnitude between branches \citep{Olson2006,Wulff2026}, we set $f_\mathrm{dip,TL} \approx 0.1 f_\mathrm{dip}$. Finally,  $\mathrm{Ro}_\mathrm{\ell, crit} = 0.12$ is the critical local Rossby number of the transition. It should be noted that $0.12$ is a conservative threshold, and in reality, other factors like density stratification, the dominance of the Lorentz force, and diffusivities, may influence this transition \citep[e.g.][]{Zaire2022,Hidalgo2025a,Wulff2026}.

\subsection{Planetary magnetosphere and stellar pressure}
The magnetopause standoff distance $r_\mathrm{M}$ of a planet orbiting at a distance $r_\mathrm{orb}$ from the host star, used here as a proxy for the upstream magnetospheric size, is given by
\begin{equation}
    r_\mathrm{M}(r_\mathrm{orb}) = r_\mathrm{pl} \sqrt[6]{\frac{f^2_0 {B_\mathrm{p,0}^{(\omega)}}^2}{2\mu_0 p_\star (r_\mathrm{orb})}},
\end{equation}
where $f_0 = 1.16$ is the form factor, which represents the enhanced shielding effect produced by the magnetopause currents \citep{Chapman1931,Chapman1941,Griessmeier2004}, $\mu_0$ is the magnetic permeability of vacuum and $p_\star$ is the total stellar pressure. Assuming contributions from the magnetic, ram and thermal pressure, it yields
\begin{equation}
    p_\star (r_\mathrm{orb}) = \frac{B_\star^2 (r_\mathrm{orb})}{2 \mu_0} + \frac{\dot{M}_\star}{4\pi v_r r_\mathrm{orb}^2} \left[ v_\mathrm{rel}^2 + \frac{2k_\mathrm{B}}{m_\mathrm{P}} T(r_\mathrm{orb}) \right],
\end{equation}
where
\begin{equation}
            B_\star(r) = \left\{ \begin{array}{lcc} B_0 \left( \frac{R_\star}{r} \right)^3 & \mathrm{if} & r \leq 2.5 R_\star \\ 0.4 B_0 \left( \frac{R_\star}{r}\right)^2  & \mathrm{if} & r > 2.5 R_\star \end{array} \right.
\end{equation}
is the stellar magnetic field, $B_0$ is the average magnetic field at the stellar surface, $\dot{M}_\star$ is the mass-loss rate
\begin{equation}
    \dot{M}_\star = \dot{M}_\odot \left( \frac{R_\star}{R_\odot}\right)^{0.46} \left( \frac{R_\mathrm{X} L_\mathrm{bol}}{R_\mathrm{X,\odot} L_\mathrm{bol,\odot}}\right)^{0.77}
\end{equation}
based on the X-ray flux relation reported by \cite{Wood2021}, where 
\begin{equation}
            R_\mathrm{X} = \left\{ \begin{array}{lcc} R_\mathrm{X, sat} & \mathrm{if} & \mathrm{Ro} \leq \mathrm{Ro}_\mathrm{sat}^\mathrm{crit} \\ R_\mathrm{X,sat} \left(\mathrm{Ro}/\mathrm{Ro}_\mathrm{sat}^\mathrm{crit}\right)^\beta  & \mathrm{if} & \mathrm{Ro} > \mathrm{Ro}_\mathrm{sat}^\mathrm{crit} \end{array} \right. 
\end{equation}
with $R_\mathrm{X,sat} \approx 7.41 \cdot 10^{-4}$, $\mathrm{Ro}_\mathrm{sat}^\mathrm{crit} \approx 0.16$ and $\beta \approx -2.70$ \citep{Wright2011, Wright2018}. The velocity of the planet relative to the stellar wind is given by
\begin{equation}
    \mathbf{v}_\mathrm{rel}(r) = v_r(r) \mathbf{\hat{r}} + \left(v_\phi(r) - \sqrt{\frac{G M_\star}{r}} \right)\mathbf{\hat{\phi}},
\end{equation}
where the radial and azimuthal components are
\begin{equation}
    v_r(r) = \sqrt{\frac{2G M_\star}{R_\star}} \left( 1 - \frac{R_\star}{r} \right), 
\end{equation}
\begin{equation}
    v_\phi(r) = \left\{ \begin{array}{lcc} \Omega_\star r & \mathrm{if} & r \leq R_\mathrm{A} \\ \frac{\Omega_\star R_\mathrm{A}^2}{r}  & \mathrm{if} & r > R_\mathrm{A} \end{array} \right. ,
\end{equation}
where $\Omega_\star$ is the stellar rotation, and $R_\mathrm{A}$ is the Alfvén radius. Finally, $k_\mathrm{B}$ is the Boltzmann constant, $m_\mathrm{P}$ is the mass of the proton, and $T(r)$ is the temperature profile, given by the models of \cite{Johnstone2015a,Johnstone2015b}.

\begin{table}[t!]
\centering
\begin{tabular}{lcc}
\hline \hline
Parameter  & Modern Earth (ME) & Early Earth (EE) \\ \hline
$P_{\mathrm{rot,i}}$ & 24 h & 13 h$^{(a)}$ \\
$\omega_\mathrm{i}$ & $7.272 \times 10^{-5}$ rad/s & $1.343 \times 10^{-4}$ rad/s $^{(a)}$ \\
$B_\mathrm{p,0}$ & 1.0 G $^{(b)}$ & 0.7 G $^{(c)}$ \\
$\mathrm{Ro}_{\ell,0}$ & 0.09$^{(d)}$ & 0.04 \\
$\tau$ & 640 s $^{(e)}$ & 37 s \\
$Q$ & $\sim 12$ $^{(f)}$ & $\sim 100$ \\
$k_2$ & 0.30 $^{(e)}$ & 0.35 \\
\hline
\end{tabular}
\label{tab:planet_parameters}
\caption{Initial planetary parameters.\\
\textbf{Ref.} $(a)$ \cite{Eulenfeld2023}, $(b)$ \cite{Vidotto2013}, $(c)$ \cite{Tarduno2010}, $(d)$ \cite{Olson2006}, $(e)$ \cite{Barnes2017}, $(f)$ \cite{Williams1978}   }
\end{table} 

\begin{figure*}[ht!]
	\centering
	\includegraphics[width=\hsize]{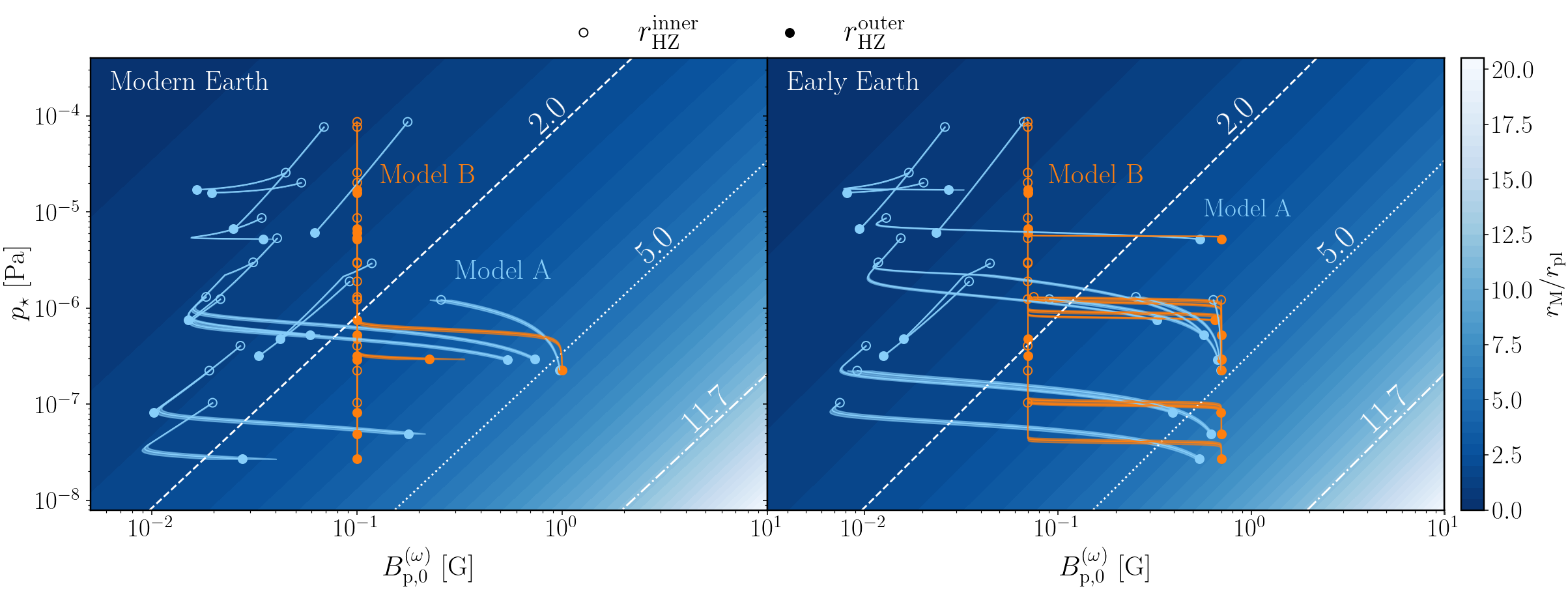}
	\caption{Normalized magnetospheric standoff distance $r_\mathrm{M}/r_\mathrm{pl}$ as a function of the planetary magnetic field and the total stellar pressure for Modern Earth (\textit{left panel}) and Early Earth (\textit{right panel}) across the HZ of the host star.}
	\label{fig:fig_wide}
\end{figure*}

\begin{figure*}[ht!]
	\centering
	\includegraphics[width=\hsize]{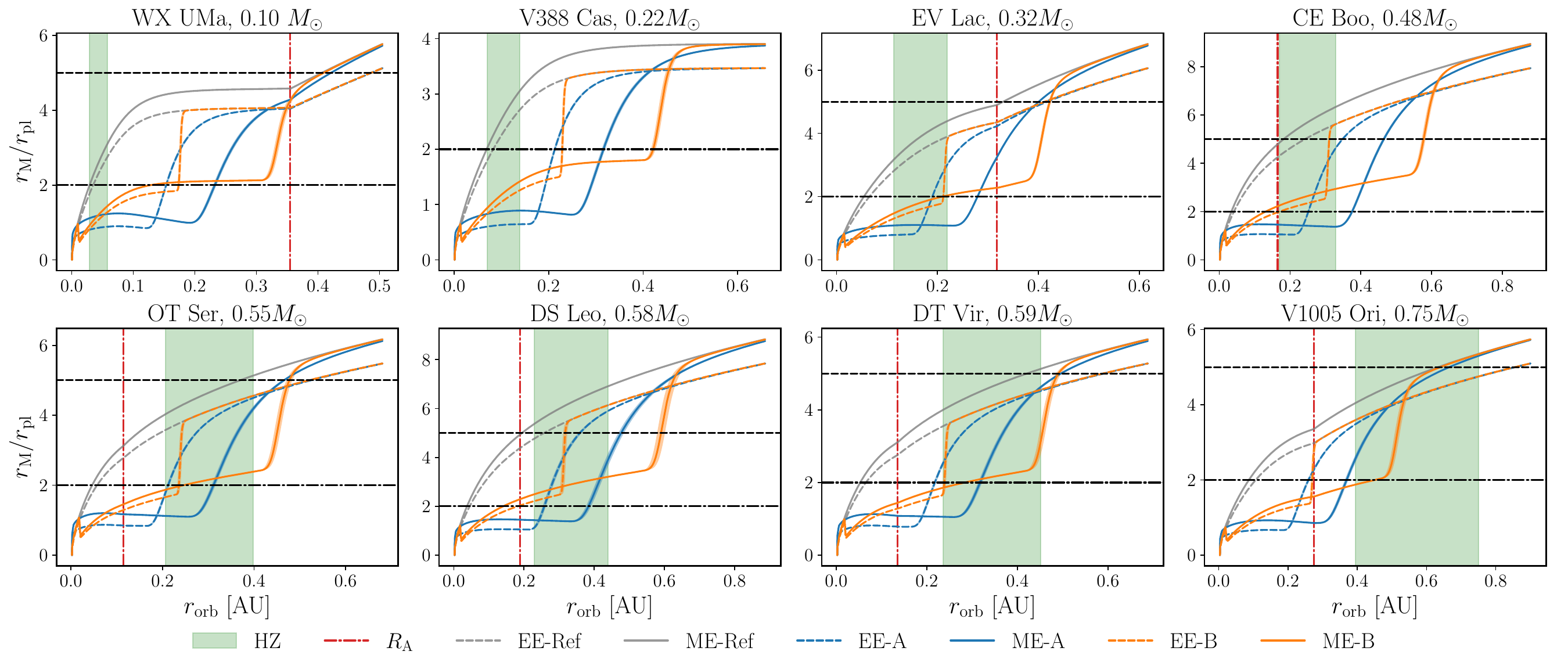}
	\caption{The size of the upstream magnetosphere as a function of the orbital distance from representative cases. Solid lines denote Modern Earth (ME) planets, while dashed lines indicate Early Earth (EE) planets. The vertical red line is the Alfvén radius $R_\mathrm{A}$, and the horizontal lines represent magnetospheres $r_\mathrm{M}/r_\mathrm{pl}$ of 2 (dashdot) and 5 (dashed).}
	\label{fig:figure2}
\end{figure*}

\subsection{Modeling Earth-like planets}
For the hypothetical planets, we adopted two distinct
baselines modeled after different epochs of Earth: a Modern Earth-like planet (ME) and an Early Earth-like planet (EE) based on the Paleoarchean. The rotational, magnetic, and tidal parameters of each model are summarized in Table~1. For both models, we adopt standard Earth structural parameters, i.e. $r_\mathrm{pl} = 6378~\mathrm{km}$, $M_\mathrm{pl} = 5.972 \cdot 10^{24}~\mathrm{kg}$ and $\alpha=0.3$ \citep{Griessmeier2009}. In the EE model, we scale the initial local Rossby number to $\mathrm{Ro}_\mathrm{\ell, 0} = 0.04$, as a consequence of the faster rotation rate. Additionally, as Earth’s tidal dissipation factor $Q$ was likely larger in the geological past \citep{Green2017}, we set an initial value of $Q \approx 100$, which is also consistent with other terrestrial bodies, such as Io and Mars \citep{Seagatz1988, Pou2022}. To estimate the constant time lag of EE, we follow $\tau \approx (2|\omega - n|Q)^{-1} \approx (2\omega_\mathrm{i} Q)^{-1}$, which yields $\tau \approx 37~\mathrm{s}$. Finally, we adopt $k_2 \approx 0.35$ to account for a hotter interior prior to the crystallization of the inner core.

\begin{figure*}[ht!]
	\centering
	\includegraphics[width=\hsize]{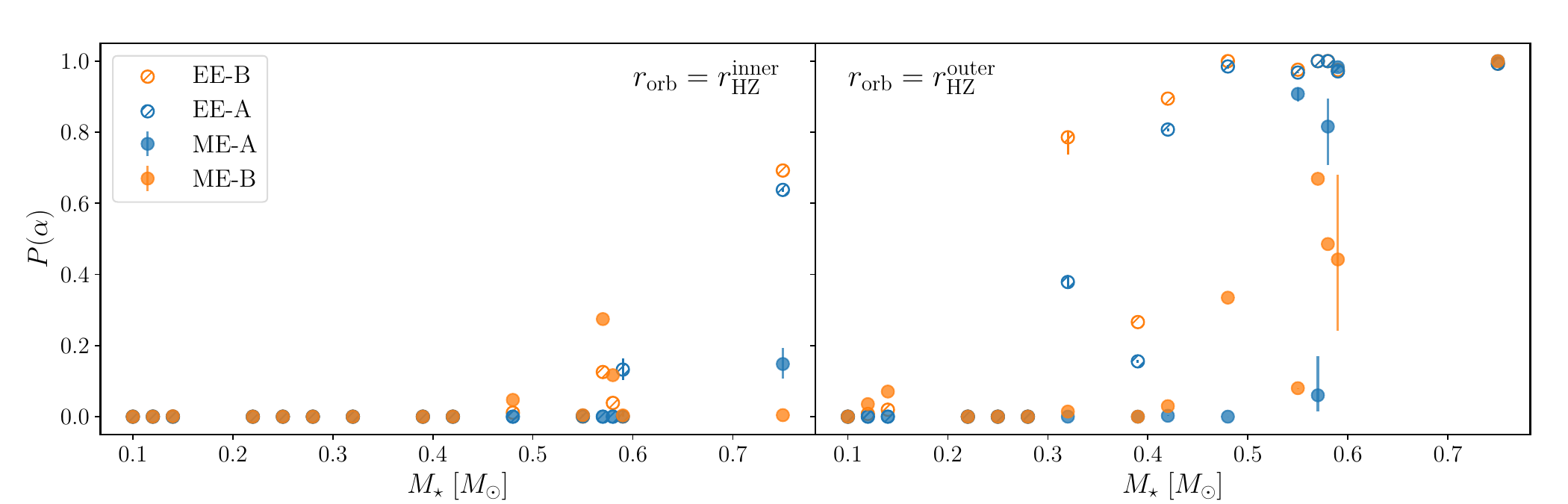}
	\caption{Likelihoods of atmospheric protection at the inner (\textit{left panel}) and outer (\textit{right panel}) boundaries of the HZ of our sample.}
	\label{fig:figure3}
\end{figure*}

\begin{figure*}[ht!]
	\centering
	\includegraphics[width=\hsize]{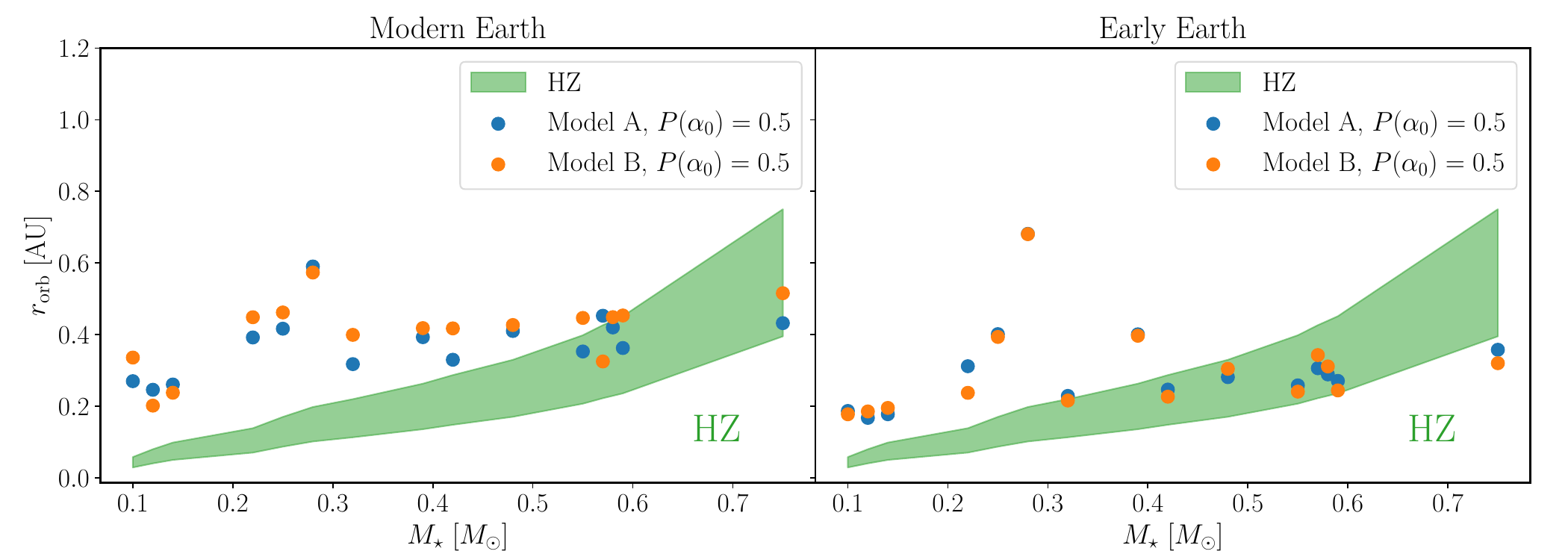}
	\caption{Required orbital distance to achieve an atmospheric protection likelihood of 0.50 as a function of stellar mass, for ME (\textit{left panel}) and EE (\textit{right panel}).}
	\label{fig:figure4}
\end{figure*}

\section{Results \label{results}}

We apply our framework to hypothetical Earth-like planets orbiting the HZs of 15 M dwarfs selected from the literature, with masses ranging between $0.10 M_\odot$ and $0.75 M_\odot$. The stellar sample is summarized in Appendix A of \cite{Hidalgo2026}, and it consists in data from \cite{Donati2008a}, \cite{Morin2010}, \cite{2008MNRAS.384...77M} and \cite{2008MNRAS.390..567M}. The HZ of these stars is estimated using the runaway greenhouse and maximum greenhouse limits from the models of \cite{Kopparapu2013, Kopparapu2014}, for the inner $r_\mathrm{HZ}^\mathrm{inner}$ and outer $r_\mathrm{HZ}^\mathrm{outer}$ radii, respectively.

The sizes of the upstream magnetospheres of ME and EE planets, under models A and B, across the HZs of the stellar sample are shown in Figure~\ref{fig:fig_wide}. Additionally, the magnetospheric sizes as a function of orbital distance for representative cases are shown in Figure~\ref{fig:figure2}. It is visible that tidal locking significantly reduces the planetary magnetic fields in most of the cases. For the ME case, only a planet orbiting V1005 Ori at $r_\mathrm{HZ}^\mathrm{outer}$ retains the baseline field of $\sim 1$,G, resulting in a magnetospheric standoff distance of $r_\mathrm{M}/r_\mathrm{pl} \approx 5.3$ under both models. In the remaining cases the magnetospheres are significantly affected, being $r_\mathrm{M}/r_\mathrm{pl} < 2$ in most of the sample, especially at $r_\mathrm{HZ}^\mathrm{inner}$. In the EE case, the situation at $r_\mathrm{HZ}^\mathrm{inner}$ is similar, but conditions at $r_\mathrm{HZ}^\mathrm{outer}$ are considerably more favorable than for ME, with a larger fraction of planets maintaining $r_\mathrm{M}/r_\mathrm{pl} > 2$.

\subsection{Atmospheric protection likelihood}

Following \cite{RodriguezMozos2019}, we define the atmospheric protection likelihood $P(\alpha_0)$ based on the Gaussian decay of the unprotected polar cap angle $\alpha_0 = \arcsin\left(  (r_\mathrm{M}/r_\mathrm{pl})^{-1/2}\right)$

\begin{equation}
    P(\alpha_0) = \left\{ \begin{array}{lcc} 1 & \mathrm{if} & \alpha_0 \leq \alpha_\mathrm{0,crit} \\ \exp\left(-\frac{1}{2}  \left[ \frac{\alpha_0 - \alpha_\mathrm{0,crit}}{\sigma} \right]^2 \right)  & \mathrm{if} & \alpha_0 > \alpha_\mathrm{0,crit} \end{array} \right. ,
\end{equation}
where $r_\mathrm{M}/r_\mathrm{pl} \geq 5$ ($\alpha_\mathrm{0,crit} \approx 26.6^\circ$) yields likelihoods of 1.0, and $\sigma$ is calibrated such
that the likelihood is 0.0 when $r_\mathrm{M}/r_\mathrm{pl} \leq 2$ ($\alpha \approx 45^\circ$), as proposed by \cite{Lammer2007}.

In Figure~\ref{fig:figure3} the atmospheric protection likelihoods of ME and EE planets orbiting the HZs of our sample are displayed. At $r_\mathrm{HZ}^\mathrm{inner}$ (left panel) the protection likelihoods are essentially zero across nearly the entire sample, regardless of the chosen model. At the outer boundary of the HZ $r_\mathrm{HZ}^\mathrm{outer}$ (right panel), the EE model exhibits a clear mass-dependent threshold at around $M_\star = 0.32 M_\odot$. In lower-mass hosts, the planets experience a rapid tidal synchronization, decreasing the planetary magnetic field significantly. Additionally, in these stars the HZ is typically in the sub-Alfvénic regime, which also significantly reduces the likelihoods. The most massive stars of the sample ($M_\star \geq 0.48 M_\odot$) offer robust protection, achieving likelihoods of $97-100\%$. On the other hand, the ME scenario lacks a clear mass-dependent transition, as tidal locking significantly reduces the magnetic fields across almost the entire sample. However, it is possible to find planets with $P(\alpha_0) \geq 0.5$ only in hosts with $M_\star > 0.50 M_\odot$.

The mentioned trends are visible in Figure~\ref{fig:figure4}, where the required orbital distances to achieve protection likelihoods of 0.50 are shown. In both cases, the required distances overlap with the HZ only in the most massive hosts of the sample.

\section{Conclusions \label{conclusions}}

We find that tidal braking systematically weakens the planetary dipolar magnetic field, substantially reducing magnetic shielding throughout the habitable zones of M dwarfs. As a result, our predictions are considerably more pessimistic than studies assuming freely rotating Earth-like planets \citep[e.g.,][]{Vidotto2013, RodriguezMozos2019}. Planets orbiting at the inner HZ are almost entirely tidally locked. Consequently, their dynamos are either significantly weakened (Model A) or driven into a multipolar regime (Model B) for both EE and ME scenarios, leaving their atmospheres highly vulnerable to stellar wind erosion. These effects are particularly severe around low-mass M dwarfs ($0.093-0.39~M_\odot$), whose habitable zones lie predominantly within the sub-Alfvénic regime, exposing planets to high stellar wind pressures.

At the outer HZ boundary, the EE scenario exhibits a clear transition at $M_\star \approx 0.32~M_\odot$. Below this threshold, tidal locking suppresses the dipolar field and the atmospheric protection likelihood is essentially zero. Above it, EE planets typically avoid rotational synchronization and remain in the super-Alfvénic regime, preserving strong dipolar fields and protection likelihoods of $97-100\%$. In contrast, ME planets generally experience weaker magnetic shielding because of their slower rotation. Overall, our results indicate that a shielded environment around M dwarfs requires a host star massive enough to place the HZ at sufficiently wide orbital distances, thereby avoiding the sub-Alfvénic regime, while planetary tidal dissipation must remain low enough to prevent rapid rotational synchronization.

Standard analytical frameworks inevitably simplify complex magnetohydrodynamic (MHD) processes like planetary dynamos and their interaction with stellar winds. While scaling laws are widely used to model planetary magnetic fields \citep[e.g.,][]{Christensen2010, 2013GeoJI.195...67D, RodriguezMozos2019}, dynamos are typically driven by non-linear fluid dynamics governed by turbulent convection and rotational shear \citep{Ortiz2023, Kapyla2023, Hidalgo2024}. Extending this work with self-consistent 3D MHD simulations will therefore be an important next step toward capturing the interaction between stellar winds and the complex multipolar magnetic fields of tidally locked planets.

\section*{Acknowledgments}
{JPH acknowledges financial support from ANID/DOCTORADO BECAS CHILE 72240057. DRGS
gratefully acknowledges support by the ANID BASAL project FB21003 and via
the Alexander von Humboldt - Foundation, Bonn, Germany.}

\bibliographystyle{cs23proc}
\bibliography{example.bib}

\end{document}